\documentclass[english,twocolumn,pra,superscriptaddress]{revtex4}
\usepackage{amssymb}
\usepackage{amsmath}
\usepackage{bbm}
\usepackage{epsfig}
\usepackage{color}
\usepackage{multirow}
\usepackage{booktabs}
\usepackage{dsfont}
\usepackage{tikz}
\usepackage{appendix}
\usepackage[colorlinks,
            linkcolor=red,
            anchorcolor=green,
            citecolor=blue
            ]{hyperref}
\begin{document}
\title{Dynamically characterizing topological phases by high-order topological charges}
\author{Wei Jia}
\affiliation{International Center for Quantum Materials and School of Physics, Peking University, Beijing 100871, China}
\affiliation{Collaborative Innovation Center of Quantum Matter, Beijing 100871, China}
\author{Lin Zhang}
\affiliation{International Center for Quantum Materials and School of Physics, Peking University, Beijing 100871, China}
\affiliation{Collaborative Innovation Center of Quantum Matter, Beijing 100871, China}
\author{Long Zhang}
\affiliation{International Center for Quantum Materials and School of Physics, Peking University, Beijing 100871, China}
\affiliation{Collaborative Innovation Center of Quantum Matter, Beijing 100871, China}
\author{Xiong-Jun Liu}
\thanks{Correspondence addressed to: xiongjunliu@pku.edu.cn}
\affiliation{International Center for Quantum Materials and School of Physics, Peking University, Beijing 100871, China}
\affiliation{Collaborative Innovation Center of Quantum Matter, Beijing 100871, China}
\affiliation{CAS Center for Excellence in Topological Quantum Computation, University of Chinese Academy of Sciences, Beijing 100190, China}
\affiliation{Institute for Quantum Science and Engineering and Department of Physics, Southern University of Science and Technology, Shenzhen 518055, China}

\begin{abstract}
We propose a new theory to characterize equilibrium topological phase with non-equilibrium quantum dynamics by introducing the concept of high-order topological charges, with novel phenomena being predicted. Through a dimension reduction approach, we can characterize a $d$-dimensional ($d$D) integer-invariant topological phase with lower-dimensional topological number quantified by high-order topological charges, of which the $s$th-order topological charges denote the monopoles confined on the $(s-1)$th-order band inversion surfaces (BISs) that are $(d-s+1)$D momentum subspaces. The bulk topology is determined by the $s$th order topological charges enclosed by the $s$th-order BISs. By quenching the system from trivial phase to topological regime, we show that the bulk topology of post-quench Hamiltonian can be detected through a high-order dynamical bulk-surface correspondence, in which both the high-order topological charges and high-order BISs are identified from quench dynamics. This characterization theory has essential advantages in two aspects. First, the highest ($d$th) order topological charges are characterized by only discrete signs of spin-polarization in zero dimension (i.e. the $0$th Chern numbers), whose measurement is much easier than the $1$st-order topological charges that are characterized by the continuous charge-related spin texture in higher dimensional space.
Secondly, a more striking result is that a first-order high integer-valued topological charge always reduces to multiple highest-order topological charges with {\em unit} charge value, and the latter can be readily detected in experiment. The two fundamental features greatly simplify the characterization and detection of the topological charges and also topological phases, which shall advance the experimental studies in the near future.
\end{abstract}

\maketitle

\section{Introduction}

Topological quantum phases have become a mainstream of research in various areas, including condensed matter physics~\cite{klitzing1980new,tsui1982two,hasan2010colloquium,qi2011topological,elliott2015colloquium,wen2017colloquium,kosterlitz2017nobel}, ultracold atoms~\cite{greiner2002quantum,bloch2008many,miyake2013realizing,atala2013direct,aidelsburger2013realization,liu2014realization,jotzu2014experimental,aidelsburger2015measuring,wu2016realization,goldman2016topological,lohse2018exploring,song2018observation}, and photonic systems~\cite{verbin2013observation,lu2016topological,mittal2019photonic}. In equilibrium theory a topological phase is characterized by the bulk topological invariant defined in the ground state of the system and host protected boundary modes through the bulk-boundary correspondence. Based upon the equilibrium characterization the topological phases can be detected in experiment from the bulk-boundary correspondence, e.g. by resolving the boundary modes with angle-resolved photoelectron
spectroscopy (ARPES) or transport measurements, which identify the equilibrium topological features~\cite{konig2007quantum,hsieh2008topological,xia2009observation}. The great success has been achieved in discovering the topological matter, such as topological insulators~\cite{fu2007topological,fu2007topologicalPRB,fu2011topological,chang2013experimental}, topological semimetals~\cite{burkov2011weyl,young2012dirac,xu2015discovery,lv2015experimental}, and topological superconductors~\cite{qi2009time,bernevig2013topological,ando2015topological,sato2017topological,zhang2018observation}.

Recently, as a momentum-space counterpart of the bulk-boundary correspondence, a dynamical bulk-surface correspondence was proposed for generic $d$-dimensional ($d$D) topological phases with integer invariants, and connects the bulk topology of such equilibrium topological phase and nontrivial dynamical pattern of quench-induced quantum dynamics emerging on the so-called band inversion surfaces (BISs)~\cite{zhang2018dynamical,zhang2019characterizing}. The BISs are $(d-1)$D interfaces in Brillouin zone (BZ) where the band inversion occurs, and are characterized by that the coupling between momentum and one (pseudo)spin component in the Hamiltonian vanishes~\cite{zhang2018dynamical}. By suddenly tuning the system from initially trivial phase to topological regime, the induced quench dynamics exhibit novel dynamical topological pattern on the $(d-1)$D BISs, which is uniquely related to the bulk topology of the $d$D equilibrium phase of the post-quench Hamiltonian. The dynamical bulk-surface correspondence establishes a universal correspondence between the equilibrium topological phases and far-from-equilibrium quantum dynamics. It provides conceptually new schemes to characterize and detect with high precision the equilibrium topological phases via non-equilibrium quench dynamics, which have been widely studied in the recent experiments with ultracold atoms~\cite{sun2018highly,sun2018uncover,yi2019observing,song2019observation,wang2020realization}, solid state spin systems~\cite{wang2019experimental,ji2020quantum,xin2020experimental}, and superconducting circuits~\cite{niu2020simulation}.
Many novel issues have been further investigated in theory, such as the dynamical characterization of both symmetry-breaking order and topological phases in correlated systems~\cite{zhang2019emergent}, the topological phases in non-Hermitian systems~\cite{zhou2018non,qiu2019fixed,wang2019simulating,zhu2020dynamic} and Floquet bands~\cite{zhang2020unified,hu2020dynamical}, generalization to generic quenches from a trivial or nontrivial phase via loop unitary construction~\cite{hu2020topological}, and to the regime with slow nonadiabatic quantum quenches~\cite{ye2020emergent}. These studies also benefit from the high controllability of the synthetic quantum systems, which facilitates the exploration of non-equilibrium quantum dynamics~\cite{caio2015quantum,hu2016dynamical,wang2017scheme,flaschner2018observation,qiu2018dynamical,mcginley2019classification,yang2019topological,tian2019observation,xiong2019nonanalyticity,chen2020linking,lu2020ideal,su2020quench}. 

The topology emerging on BISs can also be characterized by the topological charges enclosed in the BISs~\cite{zhang2019dynamical}, as an analogy to the Gaussian theorem, and such topological charges are dual to BISs and denote the monopole charges located at the nodes of the (pseudo)spin-orbit (SO) couplings~\cite{wang2019experimental,ji2020quantum,yi2019observing}. In this picture the topological invariant is viewed as the quantized flux of the monopoles through the BISs, which provides an intuitive perspective for the nontrivial bulk topology.
More recently, the high-order BISs are proposed based on a dimension reduction approach~\cite{yu2020highorder}, and the $s$th-order BIS correspond to $(d-s)$D momentum subspace which is reduced from $(s-1)$-order BIS by further taking the coupling between momentum and the $s$th (pseudo)spin component to be zero. In the quench dynamics the equilibrium topological phase can be characterized by the dynamical topology emerging on arbitrary high-order BISs. Since the higher-order BISs can be determined with less information, the dynamical theory based on high-order BISs can simplify the characterization of topological phases~\cite{yu2020highorder}. The concept of high-order BIS is novelly extended by Li etal~\cite{li2020topological} to characterize the high-order topological phases~\cite{benalcazar2017quantized,benalcazar2017electric,schindler2018higher}.
An interesting consideration is to extend the topological charges to the high-order regime based on the dimension reduction approach, which are dual to the high-order BISs and may have exceptional features and advantages in the dynamical characterization of topological phases, but have yet to be studied.

In this article, we introduce the concept of high-order topological charges, with which we propose a new dynamical characterization theory of topological phases. The equilibrium bulk topology is generically determined by the total $s$th-order topological charges confined on the $(s-1)$th-order BISs and enclosed in the $s$th-order BISs. By quenching the system from trivial phase to topological regime, we further show that the topological phase of post-quench Hamiltonian can be detected through a high-order dynamical bulk-surface correspondence, in which both the high-order topological charges and high-order BISs are identified from quench dynamics. The proposed new characterization theory has two essential advantages: (i) Unlike the $1$st-order topological charge whose characterization necessitates to measure the continuous charge-related (pseudo)spin texture in $d$D space, which could be tedious, the highest ($d$th) order topological charges are characterized by only discrete signs of spin-polarization in the zero dimension.
(ii) A high integer-valued topological charge of the first order always reduces to multiple highest-order topological charges with unit charge value. Then the high integer-valued topological invariant can be read by the summation of the highest-order topological charges enclosed by the highest-order BISs. The two fundamental features greatly simplify the characterization and detection the equilibrium topological phases.
Finally, these advantages of the dynamical characterization are illustrated with concrete examples.

The remaining part of this paper is organized as follows.
In Sec.~\uppercase\expandafter{\romannumeral2}, we introduce the generic theory of the new dynamical characterization.
In Sec.~\uppercase\expandafter{\romannumeral3}, our dynamical scheme is applied to two realistic models.
In Sec.~\uppercase\expandafter{\romannumeral4}, we show the decomposition of high integer-valued topological charges.
Finally, we summarize the main results and provide the brief discussion in Sec.~\uppercase\expandafter{\romannumeral5}.

\section{Generic Theory}

\subsection{Model Hamiltonian and dimension reduction}

We start with the basic Hamiltonian describing a $d$-dimensional ($d$D) gapped topological phase with integer invariant, which can be written in the elementary representation matrices of Clifford algebra~\cite{morimoto2013topological,chiu2016classification} as
\begin{equation}\label{e1}
\mathcal{H}(\mathbf{k})=\mathbf{h}(\mathbf{k})\cdot \boldsymbol{\gamma}
=\sum^{d}_{i=0}h_i(\mathbf{k})\gamma_i,
\end{equation}
where the vector field $\mathbf{h}(\mathbf{k})$ describes a $(d+1)$D Zeeman field depending on the Bloch momentum $\mathbf{k}$ in BZ.
The $\boldsymbol{\gamma}$ matrices obey anti-commutation relations $\{\gamma_i,\gamma_j\}=2\delta_{i,j}\mathbf{1}$ for $i,j=0,1,\cdots,d$, and their dimensionality is $2^{(d+1)/2}$ (or $2^{d/2}$) if $d$ is odd (or even), which is the minimal requirement to open a topological gap for the $d$D topological phase.
For 1D/2D case~\cite{su1980soliton,haldane1988model,chiu2013classification}, the $\boldsymbol{\gamma}$ matrices simply reduce to the Pauli matrices. For 3D/4D system~\cite{zhang2001four,schnyder2008classification}, the $\boldsymbol{\gamma}$ matrices are constructed as the tensor product of the Pauli matrices. The topology of this basic Hamiltonian is characterized by the $d$D (or $d/2$-th) winding number (or Chern number) if $d$ is odd (or even), which counts the coverage times of the mapping $\hat{\mathbf{h}}(\mathbf{k})=\mathbf{h}(\mathbf{k})/|\mathbf{h}(\mathbf{k})|$ from the BZ torus $T^d$ to the $d$D spherical surface $S^d$~\cite{fruchart2013introduction}.

Now we perform dimension reduction for the above Hamiltonian, and bulk topology will be reduced into the lower-dimensional subsystem. One can choose an arbitrary $h$-component, say $h_0(\mathbf{k})$, to characterize the {\em dispersion} of the decoupled bands of $\gamma_0$. Accordingly, the remaining $h$-components are denoted as the SO vector field $\mathbf{h}_{\rm so}(\mathbf{k})\equiv(h_1(\mathbf{k}),h_2(\mathbf{k}),\dots,h_d(\mathbf{k}))$. The SO vector field opens a topological gap at the band-crossing with $h_0(\mathbf{k})=0$, which is defined as the (first-order) BISs, namely $\mathcal{B}_1\equiv\{\mathbf{k}\in\mathrm{BZ}\vert h_0(\mathbf{k})=0\}$. The bulk-surface duality has manifested that the bulk topology can be reduced to the winding of the $d$D SO vector field on the $(d-1)$D first-order BISs~\cite{zhang2018dynamical}. This lower-dimensional topology can be treated in an effective $(d-1)$D gapped Hamiltonian on the first-order BISs,
\begin{equation}\label{eh2}
\mathcal{H}^{(1)}_{\rm eff}(\tilde{\mathbf{k}})=\mathbf{h}_{\rm so}(\tilde{\mathbf{k}})\cdot\tilde{\boldsymbol{\gamma}}=\sum_{i=1}^{d}h_i(\tilde{\mathbf{k}})\tilde{\gamma}_i,\qquad\tilde{\mathbf{k}}\in\mathcal{B}_1,
\end{equation}
where $\tilde{\boldsymbol{\gamma}}$ are the corresponding gamma matrices on the $(d-1)$D subspace. The topological number of Hamiltonian (\ref{eh2}) is given by
the coverage times of the mapping $\hat{\mathbf{h}}_{\rm so}(\tilde{\mathbf{k}})=\mathbf{h}_{\rm so}(\tilde{\mathbf{k}})/|\mathbf{h}_{\rm so}(\tilde{\mathbf{k}})|$ from $\mathcal{B}_1$ to $S^{(d-1)}$. Now we can also define BISs for $\mathcal{H}^{(1)}_{\rm eff}$, which is called the second-order BISs~\cite{yu2020highorder}. Without loss of generality, the component $h_1(\tilde{\mathbf{k}})$ is used to define the $(d-2)$D second-order BISs as $\mathcal{B}_2\equiv\{\tilde{\mathbf{k}}\in\mathcal{B}_1\vert h_1(\tilde{\mathbf{k}})=0\}=\{\mathbf{k}\in\mathrm{BZ}\vert h_0(\mathbf{k})=h_1(\mathbf{k})=0\}$. Then the bulk topology is reduced to the winding of the $(d-1)$D effective SO vector field $\mathbf{h}^{(1)}_{\rm so}(\tilde{\mathbf{k}})\equiv(h_2(\tilde{\mathbf{k}}),\dots,h_d(\tilde{\mathbf{k}}))$ on the second-order BISs.

By repeating the above dimension reduction procedure, the $d$D bulk topology can be reduced to the integer invariant of $(d-s+1)$D effective Hamiltonian on the $(s-1)$th-order BISs $\mathcal{B}_{s-1}=\{\mathbf{k}\in{\rm BZ}\vert h_{0}(\mathbf{k})=\cdots=h_{s-2}(\mathbf{k})=0\}$,
\begin{equation}\label{e2}
    \mathcal{H}^{(s-1)}_{\rm eff}(\tilde{\mathbf{k}})=h_{s-1}(\tilde{\mathbf{k}})\tilde{\gamma}_{s-1}+\sum_{i=s}^{d}h_i(\tilde{\mathbf{k}})\tilde{\gamma}_i,\quad\tilde{\mathbf{k}}\in\mathcal{B}_{s-1},
\end{equation}
where $\tilde{\boldsymbol{\gamma}}$ are the corresponding Gamma matrices on the $(d-s+1)$D subspace. Thus $h_{s-1}(\tilde{\mathbf{k}})$ component further defines the $(d-s)$D $s$th-order BISs as
\begin{equation}
\begin{split}
\mathcal{B}_{s}& \equiv\{\tilde{\mathbf{k}}\in\mathcal{B}_{s-1}\vert h_{s-1}(\tilde{\mathbf{k}})=0\}\\
&=\{\mathbf{k}\in{\rm BZ}\vert h_{0}(\mathbf{k})=\cdots=h_{s-1}(\mathbf{k})=0\}
\end{split}
\end{equation}
for $\mathcal{H}^{(s-1)}_{\rm eff}$ [see Fig.~\ref{Fig1}(a)],
and the remaining components represent the corresponding $(d-s+1)$D effective SO vector field $ \mathbf{h}^{(s-1)}_{\rm so}(\tilde{\mathbf{k}})\equiv(h_{s}(\tilde{\mathbf{k}}),\dots,h_{d}(\tilde{\mathbf{k}}))$. The topological number is given by the winding of the $(d-s+1)$D SO vector field on the $s$th-order BISs [see Fig.~\ref{Fig1}(b)].

\subsection{High-order topological charges}

As an analogy to the Gaussian theorem, the bulk topology can also be characterized by the topological charges enclosed in the BISs, and such topological charges are dual to BISs and denote the monopole charges located at the nodes of the SO couplings~\cite{zhang2018dynamical,zhang2019characterizing,zhang2019dynamical,wang2019experimental,ji2020quantum,yi2019observing}.
In this picture the topological invariant of $\mathcal{H}^{(s-1)}_{\rm eff}$ is simply viewed as the quantized flux of the monopoles through the $s$th-order BISs.

\begin{figure}[!htbp]
\centering
\rotatebox{0}{\resizebox {8.0cm}{6.1cm} {\includegraphics{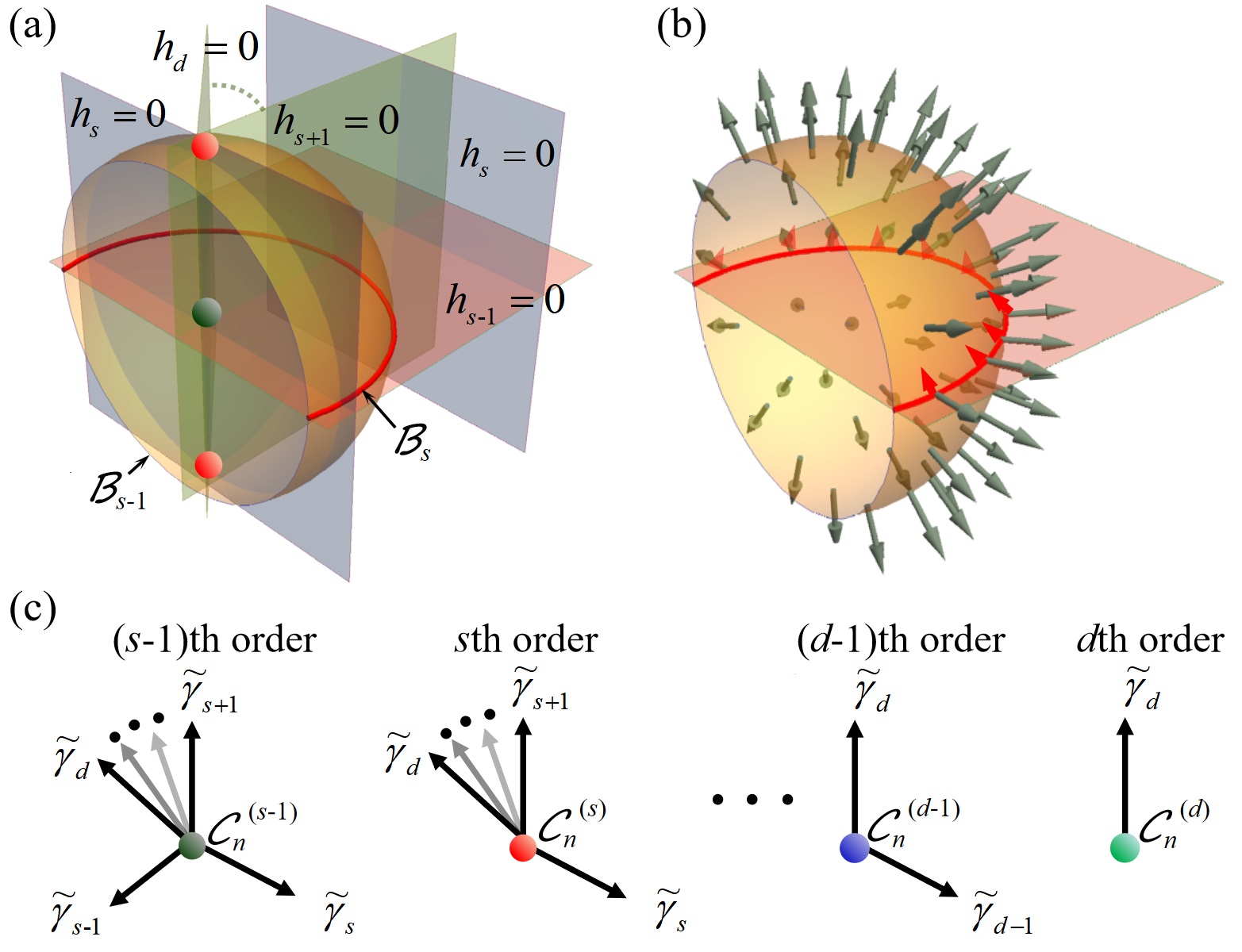}}}
\caption{(a)~A $s$th-order BIS (red curve) produced by $h_{s-1}=0$ and two $s$th-order topological charges (red points) determined by $h_{s}=h_{s+1}=\cdots=h_d=0$ are both confined on the $(s-1)$th-order BIS (hemisphere surface), while the $(s-1)$th-order topological charge (gray-green point) determined by $h_{s-1}=h_{s}=h_{s+1}=\cdots=h_d=0$ is enclosed by the $(s-1)$th-order BIS.
(b)~The $(d-s+1)$D topology described by the winding on $(s-1)$th-order BIS (gray-green arrows) is reduced on the $(d-s)$D $s$th-order BIS (red arrows).
(c)~The properties of a high-order topological charge are characterized by constructing the coordinates $\tilde{\gamma}_{s}\text{-}\tilde{\gamma}_{s+1}\text{-}\cdots\text{-}\tilde{\gamma}_{d}$ in (pseudo)spin subspace.}
\label{Fig1}
\end{figure}

We introduce the $s$th-order topological charges
\begin{equation}
\begin{split}
\mathcal{C}^{(s)}_n=&\frac{\Gamma[(d-s+1)/2]}{2\pi^{(d-s+1)/2}}\int _{\mathcal{S}_n}
\bigg[\frac{1}{|\mathbf{h}^{(s-1)}_{\rm so}|^{d-s+1}}\\
&\sum^{d}_{j=s}(-1)^{j-1}h_j\bigg] \text{d}h_{s}\wedge\cdots\wedge \text{d}h_d,\\
\end{split}
\end{equation}
which are located at the nodes $\tilde{\mathbf{k}}=\mathfrak{g}_n$ of the $(d-s+1)$D effective SO vector field with $\mathbf{h}^{(s-1)}_{\text{so}}(\mathfrak{g}_n)=0$ and characterize the corresponding monopoles.
Here $\mathcal{S}_n$ denotes a $(d-s+2)$D interface on $(s-1)$th-order BISs, enclosing the $n$th monopole $\mathfrak{g}_n$. In the typical case where $\mathbf{h}^{(s-1)}_{\rm so}$ is linear near the monopole, the $s$th-order topological charges can be simplified as
$\mathcal{C}^{(s)}_n=\text{sgn}[J_{\mathbf{h}^{(s-1)}_{\text{so}}}(\mathfrak{g}_n)]$,
where $J_{\mathbf{h}^{(s-1)}_{\text{so}}}(\tilde{\mathbf{k}})\equiv\text{det} [(\partial
h^{(s-1)}_{\text{so},j}/\partial \tilde{k}_i)]$ is Jacobian determinant with $j=s,s+1,\cdots,d$. However, when a monopole does not have the linear dispersion, the Jacobian is zero and the charge value $|\mathcal{C}^{(s)}_n|$ is in fact larger than one.

We emphasize that the $s$th-order topological charges are confined on the $(s-1)$th-order BISs $\mathcal{B}_{s-1}$ [see Fig.~\ref{Fig1}(a)] and are characterized by the all components of $(d-s+1)$D effective SO vector field $\mathbf{h}^{(s-1)}_{\rm so}(\tilde{\mathbf{k}})$. Thus the properties (charge value and chirality) of $s$th-order topological charges can be read out by measuring the (pseudo)spin structure
of $\mathbf{h}^{(s-1)}_{\rm so}(\tilde{\mathbf{k}})$ at $\tilde{\mathbf{k}}\to\mathfrak{g}_n$ in the (pseudo)spin subspace of $\tilde{\gamma}_{s}\text{-}\tilde{\gamma}_{s+1}\text{-}\cdots\text{-}\tilde{\gamma}_{d}$ coordinate system [see Fig.~{\ref{Fig1}}(c)]. In particular, one can find that the highest ($d$th) order topological charges are only characterized by the discrete signs of $h_d(\tilde{\mathbf{k}})$ at $\tilde{\mathbf{k}}\to\mathfrak{g}_n$, i.e. the $0$th Chern numbers~\cite{yu2020highorder}. This intrinsic property determines that whose charge value is only $|\mathcal{C}^{(d)}_{n}|=1$.
Moreover, a high integer-valued topological charge can always reduce to multiple highest-order topological charges with unit charge value by the dimension reduction procedure.
This two fundamental features of highest-order topological charges can greatly simplify the characterization and detection the equilibrium topological phases, which avoids the redundant measurements of the continuous charge-related (pseudo)spin texture in high dimensional space and provides the easy measurements in experiments (See sections \ref{section3} and \ref{section4} for details). This is one of the key ideas of this paper.

Besides, three points are worthwhile to mention:
(i) The order of topological charge is actually the number of dimension reduction for bulk Hamiltonian.
(ii) The real dimensionality for the arbitrary high-order topological charge is zero, because the topological charges are the nodes of effective SO vector field in momentum subspace.
(iii) The configurations of high-order BISs are sharply different for choosing different $h$-components of the Hamiltonian, thus the location of the corresponding high-order topological charges should be different. Nevertheless, this does not affect the results of topological characterization (see Appendix \ref{appendix-4}).

\subsection{High-order dynamical bulk-surface correspondence}

We further propose to use quench dynamics to detect the high-order topological charges and the corresponding high-order BISs, which establishes the high-order dynamical bulk-surface correspondence to characterize the equilibrium topological phases.
We consider a series of deep quench process (see Appendix~\ref{appendix-2}) along all axes $\gamma_i$ with $i=0,1,\dots,d$ while only measure a single (pseudo)spin component $\gamma_0$, which is well measurable in cold atom experiments~\cite{sun2018highly,sun2018uncover,zhang2019characterizing}. Then the time-averaged (pseudo)spin polarization (TASP) is given by
\begin{align}
    \overline{\langle\gamma_{0}(\mathbf{k})\rangle}_{i} =-h_{0}(\mathbf{k})h_{i}(\mathbf{k})/E^{2}(\mathbf{k}),\label{eb}
\end{align}
where $E(\mathbf{k})=\sqrt{\sum^d_{i=0} h^2_i}$ is the energy of the post-quenched Hamiltonian. Note that the TASP vanishes both on the momentum space with $h_0(\mathbf{k})=0$ and $h_i(\mathbf{k})=0$.

\begin{figure*}[!htbp]
\begin{center}
\rotatebox{0}{\resizebox {13.0cm}{8.5cm} {\includegraphics{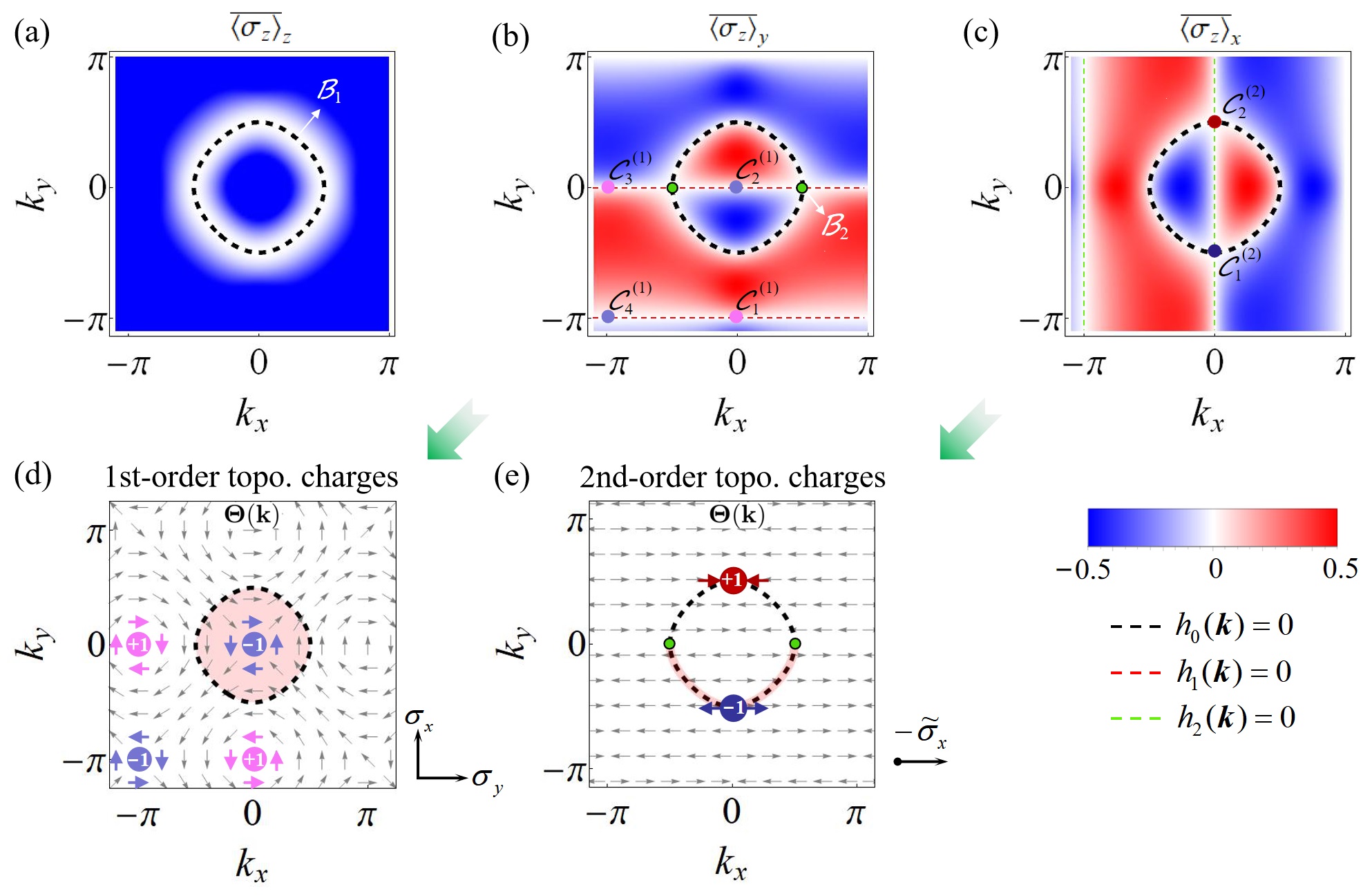}}}
\caption{Dynamical characterization of 2D QAH model.
(a)-(c)~The TASP $\overline{\langle\sigma_z(\mathbf{k}) \rangle}_{z,y,x}$ via quenching $m_{0,1,2}$ along all axes, where the vanishing polarization are marked as black, red, and green dashed lines that represent the interfaces with $h_{0,1,2}(\mathbf{k})=0$ respectively. The first-order BIS $\mathcal{B}_1$ (black dashed line) is given by $h_0(\mathbf{k})=0$ in (a). The second-order BISs $\mathcal{B}_2$ (green points) at $\mathbf{k}=(\pm\pi/2,0)$ are given by $h_1(\mathbf{k})=0$ on first-order BIS $\mathcal{B}_1$ in (b), and $h_1(\mathbf{k})=h_2(\mathbf{k})=0$ gives the first-order topological charges $\mathcal{C}_{n=1,2,3,4}^{(1)}$ (light-pink and light-blue points) at $(0,-\pi)$, $(0,0)$, $(-\pi,0)$, and $(-\pi,-\pi)$. The second-order topological charges $\mathcal{C}_1^{(2)}=-1$ (blue point) at $\mathbf{k}=(0,-\pi/2)$ and $\mathcal{C}_2^{(2)}=1$ (red point) at $\mathbf{k}=(0,\pi/2)$ are given by $h_2(\mathbf{k})=0$ on first-order BIS $\mathcal{B}_1$ in (c).
(d)~The normalized dynamic field in spin space of $\sigma_y-\sigma_x$ characterizes the properties of first-order topological charges, where $\mathcal{C}_{2}^{(1)}$ in the region $h_0(\mathbf{k})<0$ (light-red surface) gives the 1st Chern number $\text{Ch}_1=\mathcal{C}_2^{(1)}=-1$.
(e)~The normalized dynamic field in spin subspace of $\tilde{\sigma}_x$ characterizes the properties of second-order topological charges, where $\mathcal{C}_1^{(2)}=-1$ in the region $h_1(\tilde{\mathbf{k}})<0$ (light-red solid curves) gives the 1st Chern number $\text{Ch}_1=\mathcal{C}_1^{(2)}=-1$. Here the other parameter is $t_{\text{so}}=t_0$.}
\label{Fig2}
\end{center}
\end{figure*}

Now the high-order topological charges and high-order BISs can be identified by measuring TASP. We define a set
$\mathcal{S}^{(i)}\equiv\{\mathbf{k}\in\mathrm{BZ}\vert\overline{\langle\gamma_0(\mathbf{k})\rangle}_i=0, \overline{\langle\gamma_0(\mathbf{k})\rangle}_0\neq 0\} $ for $i>0$, which includes the momenta with $h_i=0$ but $h_0\neq 0$. Then the closure $\bar{\mathcal{S}}^{(i)}$ also contains the momenta with $h_i=h_0=0$.
After setting $\mathcal{S}^{(0)}=\{\mathbf{k}\in\mathrm{BZ}|\overline{\langle \gamma_0(\mathbf{k}) \rangle}_{0}=0\}$, the vanishing TASP gives the $s$th-order BISs when quenching the axes $\gamma_{0},\gamma_{1},\cdots,\gamma_{s-1}$, i.e.
\begin{equation}\label{eg1}
\mathcal{B}_s=\mathcal{S}^{(0)}\cap\bar{\mathcal{S}}^{(1)}\cap\cdots\cap\bar{\mathcal{S}}^{(s-1)}.
\end{equation}
Correspondingly, the location of the $s$th-order topological charges can be determined by the momenta $\{\mathfrak{g}_{n}\}=\mathcal{B}_{s-1}\cap\bar{\mathcal{S}}^{(s)}\cap\bar{\mathcal{S}}^{(s+1)}\cap\cdots\cap\bar{\mathcal{S}}^{(d)}$. We further define the dynamical field
\begin{equation}\label{eg2}
\Theta_j(\tilde{\mathbf{k}})\equiv -\lim_{\mathbf{k}\to\tilde{{\mathbf{k}}}}\frac{\text{sgn}[h_{s-1}({\mathbf{k}})]}{\mathcal{N}_{{\mathbf{k}}}}\frac{\overline{\langle\gamma_0({\mathbf{k}}) \rangle}_{j}\overline{\langle\gamma_0({\mathbf{k}}) \rangle}_{s-1}}{\overline{\langle\gamma_0({\mathbf{k}}) \rangle}_0}
\end{equation}
in (pseudo)spin subspace of $\tilde{\gamma}_{s}\text{-}\tilde{\gamma}_{s+1}\text{-}\cdots\text{-}\tilde{\gamma}_{d}$ coordinate system, where $\mathcal{N}_{\tilde{\mathbf{k}}}$ is a normalization factor and $j=s,s+1,\cdots,d$. Near the monopole charges, the dynamic field satisfies
\begin{equation}\label{eg3}
\Theta_j(\tilde{\mathbf{k}})|_{\tilde{\mathbf{k}}\rightarrow \mathfrak{g}_n}=h^{(s-1)}_{\text{so},j}(\tilde{\mathbf{k}}),
\end{equation}
whose (pseudo)spin structures intuitively give the properties of the $s$th-order topological charges.

Finally, the bulk topology can be read out by the total $s$th-order topological charges in the regions with $\mathcal{B}_{s,-} \equiv\{\tilde{\mathbf{k}}\in\mathcal{B}_{s-1}|h_{s-1}(\tilde{\mathbf{k}})<0\}$ enclosed by the $s$th-order BISs, namely
\begin{equation}\label{eg4}
\mathcal{W}=\sum_{n\in \mathcal{B}_{s,-}}\mathcal{C}^{(s)}_n.
\end{equation}
The results of Eqs.~(\ref{eg1})-(\ref{eg4}) manifest a high-order dynamical bulk-surface correspondence, and provide the direct measurements of bulk topology via the well-resolved TASP in experiments.
In addition, it is worth mentioning that we also provide another dynamical characterization scheme by quenching all (pseudo)spin axes and measuring multiple (pseudo)spin axis in Appendix~\ref{appendix-4}. Although the measurements of multiple (pseudo)spin components are challenging in recent experiments, this scheme is easier to determine high-order BISs and high-order topological charges, and then the equilibrium topological phase, which may has broader applications in the future.

\begin{figure*}[htbp]
\centering
\rotatebox{0}{\resizebox {18.0cm}{7.3cm} {\includegraphics{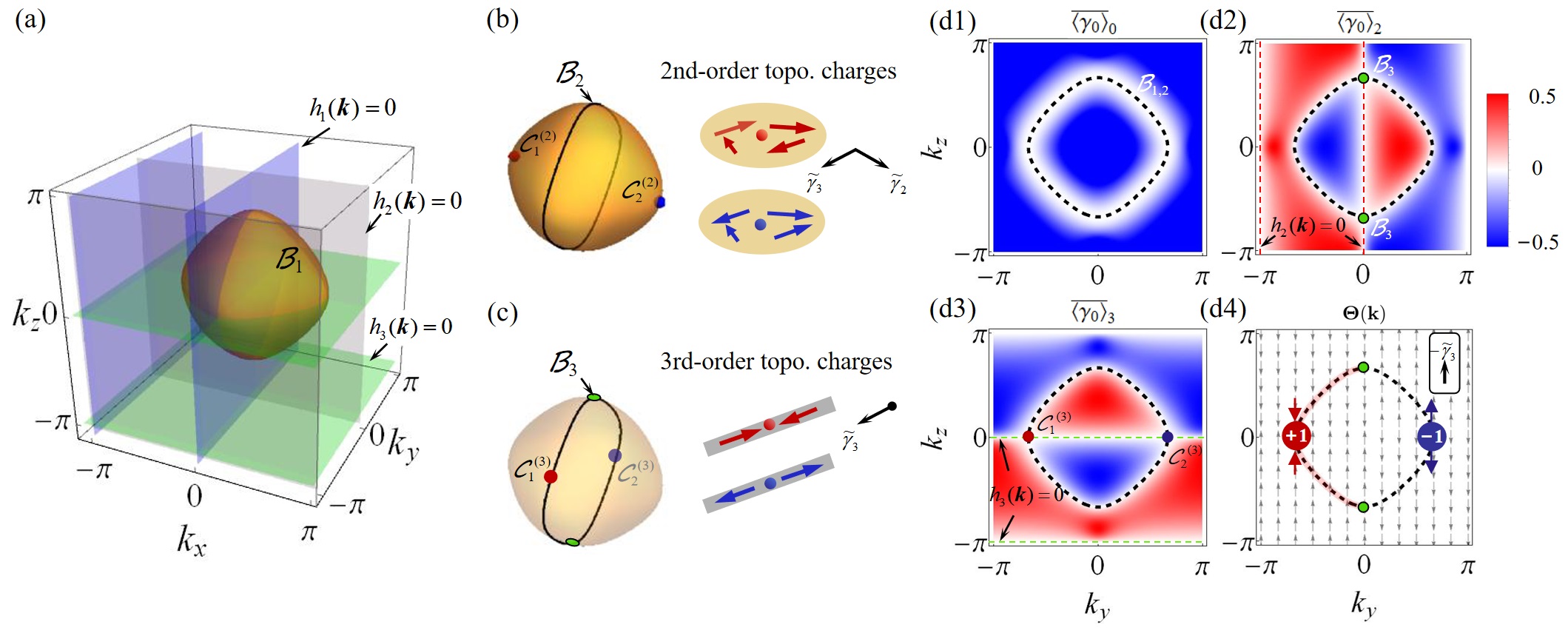}}}
\caption{Dynamical characterization of 3D chiral topological insulator. (a)~ The vanished TASP of $\overline{\langle\gamma_{0}(\mathbf{k})\rangle}_{0,1,2,3}$, where $\overline{\langle\gamma_{0}(\mathbf{k})\rangle}_{0}=0$
presents a spherical-like surface (orange surface) and gives the first-order BIS $\mathcal{B}_1$ with $h_{0}(\mathbf{k})=0$.
(b)~The second-order topological charges are determined by $h_2(\tilde{\mathbf{k}})=h_3(\tilde{\mathbf{k}})=0$ on $\mathcal{B}_1$ and are characterized by the normalized dynamic field in pseudospin subspace $\tilde{\gamma}_2-\tilde{\gamma}_3$, with $\mathcal{C}_1^{(2)}=1$ (red point) at $\mathbf{k}=(-2\pi/3,0,0)$ and $\mathcal{C}_2^{(2)}=-1$ (blue point) at $\mathbf{k}=(2\pi/3,0,0)$.
The second-order BIS $\mathcal{B}_2$ (black curve) divides $\mathcal{B}_1$ into two regions, where $\mathcal{C}_1^{(2)}=1$ in left hemisphere surface with $h_1(\tilde{\mathbf{k}})<0$ gives the winding number $\nu_3=\mathcal{C}_1^{(2)}=1$.
(c)~The third-order topological charges are determined by $h_3(\tilde{\mathbf{k}})=0$ on $\mathcal{B}_2 $ and are characterized by the normalized dynamic field in pseudospin subspace $\tilde{\gamma}_3$, with $\mathcal{C}_1^{(3)}=1$ (red point) at $\mathbf{k}=(0,-2\pi/3,0)$ and $\mathcal{C}_2^{(3)}=-1$ (blue point) at $\mathbf{k}=(0,2\pi/3,0)$.
The third-order BISs $\mathcal{B}_3$ (green points) divides $\mathcal{B}_2$ into two regions, where $\mathcal{C}_1^{(3)}=1$ in the front of ring with $h_2(\tilde{\mathbf{k}})<0$ gives the winding number $\nu_3=\mathcal{C}_1^{(3)}=1$.
(d)~Minimal measurement by detecting the TASP on the 2D plane of $k_x=0$, where the vanishing polarization marked as the black, red, and green dashed lines presents the interfaces with $h_{0(1),2,3}(\mathbf{k})=0$, respectively. The $\mathcal{B}_1$ and $\mathcal{B}_2$ are coincident in (d1). $h_2(\mathbf{k})=0$ on $\mathcal{B}_2$ gives the third-order BISs $\mathcal{B}_3$ (green points) in (d2), and $h_3(\mathbf{k})=0$ gives two third-order topological charges $\mathcal{C}_1^{(3)}=1$ (red point) and $\mathcal{C}_2^{(3)}=-1$ (blue point) in (d3). The normalized dynamic field characterizes the properties of the third-order topological charges in (d4), where the leftward $\mathcal{C}_1^{(3)}=1$ in the region $h_2(\tilde{\mathbf{k}})<0$ (light-red curves) gives the winding number $\nu_3=\mathcal{C}_1^{(3)}=1$. Here the other parameter is $t_{\text{so}}=t_0$.}
\label{Fig3}
\end{figure*}

\section{Application to the realistic models}\label{section3}

We consider a simple $d$D model with
\begin{equation}
h_0=m_0-t_0\sum^d_{i=1}\cos k_{r_i},~
h_i=m_i+t_{\text{so}}\sum^d_{i=1}\sin k_{r_i},
\end{equation}
which can be realized with recent advances. Here $\mathbf{k}=(k_{r_1},k_{r_2},...,k_{r_d})$ is the $d$D momentum, $m_0$ and $m_i$ are the effective Zeeman coupling, and $t_0$, $t_{\text{so}}$ are the nearest-neighbor spin-conserved and spin-flipped hopping coefficients, respectively.

The 2D quantum anomalous Hall (QAH) model with $(r_1,r_2)=(y,x)$ is considered first, which has been realized in cold atoms experiments~\cite{liu2014realization,wu2016realization,wang2018dirac} and widely studied~\cite{zhang2013topological,pan2016bose,liu2016chiral,poon2018semimetal,jia2019topological}.
The $\boldsymbol{\gamma}$ matrices are Pauli matrices $\gamma_{0,1,2}=\sigma_{z,y,x}$.
For $m_1=m_2=0$, the bulk topology is characterized by the 1st Chern number ($\text{Ch}_1$), where the topological phase corresponds to $0<|m_0|<2t_0$ with $\text{Ch}_1=-\text{sgn}(m_0)$, but the trivial phases are for $|m_0|\geqslant 2t_0$ with $\text{Ch}_1=0$.
We perform the quench by suddenly varying $(m_0,m_1,m_2)$ from $(30t_0,0,0)$ to $(t_0,0,0)$ for $h_0$, from $(0,30t_0,0)$ to $(t_0,0,0)$ for $h_{1}$, and from $(0,0,30t_0)$ to $(t_0,0,0)$ for $h_{2}$. The time evolution of spin polarization for $\sigma_{z}$-component only needs to be measured, which can present the second-order BISs and second-order topological charges and then gives the information of bulk topology.

A ring structure characterizes the first-order BIS $\mathcal{B}_1$ with $h_0=0$ is observed from the vanishing TASP $\overline{\langle\sigma_z(\mathbf{k})\rangle}_{z}=0$  in Fig.~\ref{Fig2}(a). The vanishing polarization $\overline{\langle\sigma_z(\mathbf{k}) \rangle}_{y}=0$ in Fig.~\ref{Fig2}(b) shows the surfaces of $h_0(\mathbf{k})=0$ and $h_1(\mathbf{k})=0$, where the second-order BISs $\mathcal{B}_2$ are given by $h_1(\mathbf{k})=0$ on the first-order BIS $\mathcal{B}_1$ and present two points when taking $\mathbf{h}^{(1)}_{\text{eff-so}}(\tilde{\mathbf{k}})=h_2$. Moreover, the second-order topological charges determined by $h_2(\tilde{\mathbf{k}})=0$ sit on the first-order BIS $\mathcal{B}_1$ and are obtained by the vanishing polarization $\overline{\langle\sigma_z(\mathbf{k}) \rangle}_{x}=0$, as shown in Fig.~\ref{Fig2}(c).
Because the effective BZ is reduced as $\{\tilde{\mathbf{k}}|h_0(\mathbf{k})=0\}$ (or say $\tilde{\mathbf{k}}\in\mathcal{B}_1$) and the bottom half-ring of first-order BIS holds $h_1(\tilde{\mathbf{k}})<0$, the second-order topological charge $\mathcal{C}_1^{(2)}=1$ is enclosed into by the second-order BISs, which gives the 1st Chern number $\text{Ch}_1=\mathcal{C}_1^{(2)}=-1$ and is shown in Fig.~\ref{Fig2}(e). Compared with the dynamical characterization by using the first-order topological charges in Fig.~\ref{Fig2}(d), the spin textures around second-order topological charges are determined by the sign of $h_2(\tilde{\mathbf{k}})$ at two sides of monopoles, which is more convenient for the experimental measurement.

\begin{figure*}[!htbp]
\begin{center}
\rotatebox{0}{\resizebox {14.1cm}{8.3cm} {\includegraphics{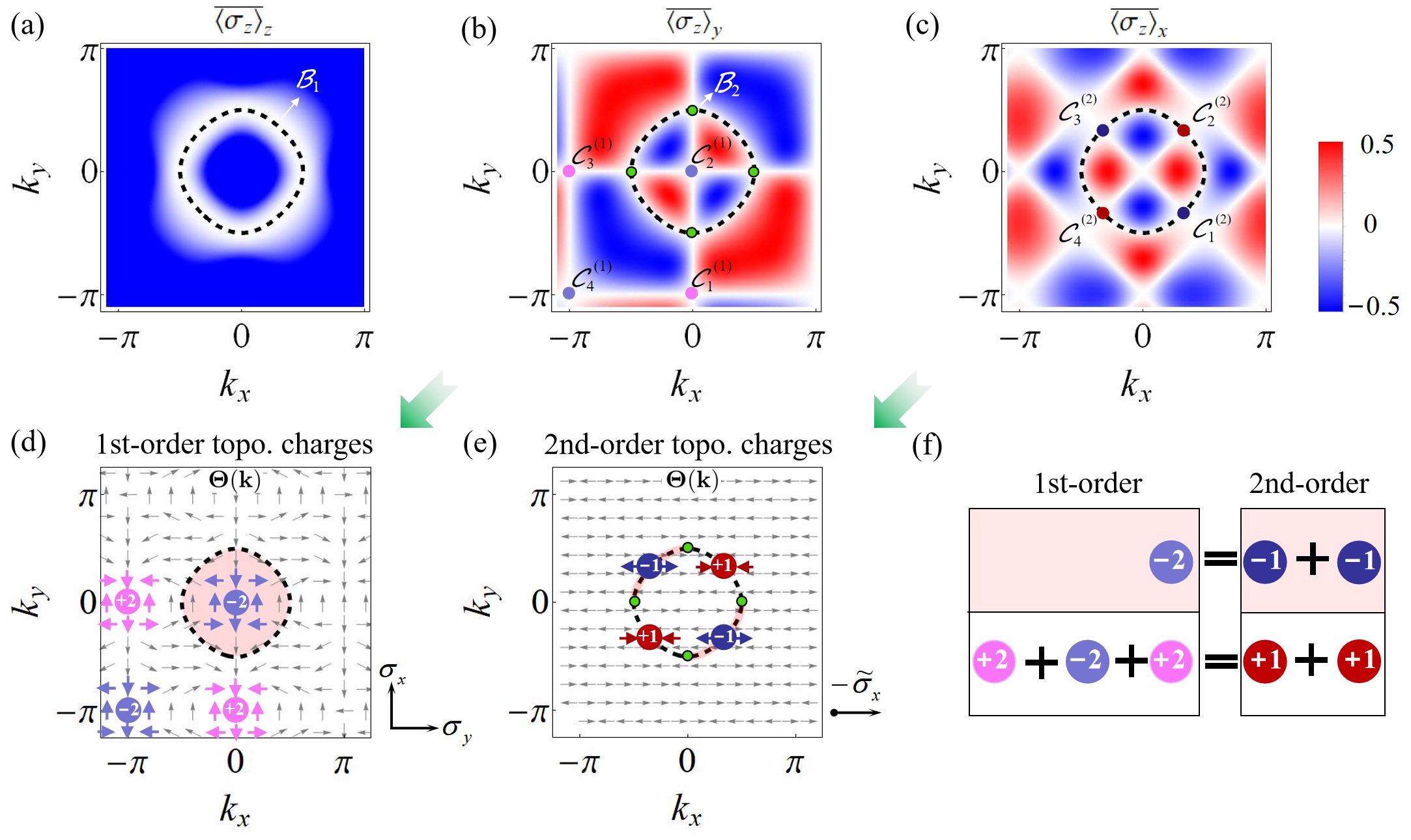}}}
\caption{Numerical results in the extended 2D QAH model with charge value $|\mathcal{C}^{(1)}_n|=2$.
(a)-(c)~The TASP $\overline{\langle\sigma_z(\mathbf{k}) \rangle}_{z,y,x}$, where the vanishing polarization presents the first-order BIS $\mathcal{B}_1$ with $h_{0}(\mathbf{k})=0$ (black dashed curve) in (a). The vanishing polarization on $\mathcal{B}_1$ gives the second-order BISs $\mathcal{B}_2$ (green points) in (b), and the vanishing polarization on $\mathcal{B}_1$ gives the second-order topological charges $\mathcal{C}^{(2)}_{n=1,2,3,4}$ (blue and red points) in (c).
(d)~The normalized dynamic field characterizes the properties of four first-order topological charges $\mathcal{C}^{(1)}_{n=1,2,3,4}$, where the first-order topological charge in the region $h_0(\mathbf{k})<0$ (light-red region) gives the 1st Chern number $\text{Ch}_1=\mathcal{C}^{(1)}_{2}=-2$.
(e)~The normalized dynamic field characterizes the properties of four second-order topological charges $\mathcal{C}^{(2)}_{n=1,2,3,4}$, where the summation of second-order topological charges in the region $h_1(\tilde{\mathbf{k}})<0$ (light-red region) gives the 1st Chern number $\text{Ch}_1=\mathcal{C}^{(2)}_{1}+\mathcal{C}^{(2)}_{3}=-2$.
(f)~The first-order topological charge $\mathcal{C}^{(1)}_{2}=-2$ enclosed by $\mathcal{B}_1$ is equivalent to the sum of two second-order topological charges $\mathcal{C}^{(2)}_{n=1,3}=-1$ enclosed by $\mathcal{B}_2$.
The sum of remaining first-order topological charges $\mathcal{C}^{(1)}_{n=1,3,4}$ is equivalent to the sum of two second-order topological charges $\mathcal{C}^{(2)}_{n=2,4}=1$.
Here the other parameter is $t_{\text{so}}=t_0$.}
\label{Fig6}
\end{center}
\end{figure*}

We further consider the application to a 3D chiral topological insulator with $(r_1,r_2,r_3)=(x,y,z)$, which has been simulated by using nitrogen-vacancy center~\cite{ji2020quantum}. The $\boldsymbol{\gamma}$ matrices are taken as $\gamma_0=\sigma_z\otimes \tau_x$, $\gamma_1=\sigma_x\otimes \mathbbm{1}$, $\gamma_2=\sigma_y\otimes \mathbbm{1}$, and $\gamma_3=\sigma_z\otimes \tau_z$, where $\sigma_{x,y,z}$ and $\tau_{x,y,z}$ are both Pauli matrices. For $m_1=m_2=m_3=0$, the topological phases are classified by the 3D winding number and are distinguished as: (i) $t_0<m_0<3t_0$ with winding number $\nu_3=1$; (ii) $-t_0<m_0<t_0$ with $\nu_3=-2$; and (iii) $-3t_0<m_0<-t_0$ with $\nu_3=1$. Beyond these regions the phase is trivial.
We perform the quench by suddenly varying $(m_0,m_1,m_2,m_3)$ from $(30t_0,0,0,0)$ to $(1.5t_0,0,0,0)$ for $h_0$, from $m_i=30t_0$ to $0$ for $h_i$
(tuning $m_0$ to $1.5t_0$ and keeping $m_{j\neq i}=0$), then the bulk topology in region (i) can be read out by measuring the time evolution of pseudospin polarization of the $\gamma_{0}$-component.

The vanishing $\overline{\langle\gamma_{0}(\mathbf{k})\rangle}_{0,1,2,3}$ in six 2D planes of BZ implies $h_{1,2,3}(\mathbf{k})=0$, but the spherical-like surface is for $h_{0}(\mathbf{k})=0$, which identifies the first-order BIS $\mathcal{B}_1$ [see Fig.~\ref{Fig3}(a)].
When taking the effective SO vector field as $\mathbf{h}^{(1)}_{\text{eff-so}}(\tilde{\mathbf{k}})=(h_2,h_3)$, the second-order BIS $\mathcal{B}_2$ presents the ring-shape structure produced by $h_{0}=h_{1}=0$ in vanishing polarization of $\overline{\langle\gamma_{0}(\mathbf{k})\rangle}_{1}$, which are confined on the first-order BIS $\mathcal{B}_1$ [see Fig.~\ref{Fig3}(b)]. The corresponding second-order topological charges $\mathcal{C}^{(2)}_{n=1,2}$ with $h_2=h_3=0$ are determined by the vanishing polarization of $\overline{\langle\gamma_{0}(\mathbf{k})\rangle}_{2,3}$ on the first-order BIS $\mathcal{B}_1$. The second-order topological charge $\mathcal{C}_1^{2}=1$ is enclosed by the second-order BIS $\mathcal{B}_2$, giving the 3D winding number $\nu_3=\mathcal{C}_1^{(2)}=1$. Moreover, when the effective SO vector field $\mathbf{h}^{(3)}_{\text{eff-so}}(\tilde{\mathbf{k}})=h_3$ is taken, the third-order BISs $\mathcal{B}_3$ are produced by $h_0=h_1=h_2=0$ in vanishing polarization of $\overline{\langle\gamma_{0}(\mathbf{k})\rangle}_{0,1,2}$, which are confined on the second-order BIS $\mathcal{B}_2$ and present two points [see Fig.~\ref{Fig3}(c)]. The effective BZ is reduced as $\{\tilde{\mathbf{k}}|h_0(\mathbf{k})=h_1(\mathbf{k})=0\}$ (or say $\tilde{\mathbf{k}}\in \mathcal{B}_2$) and the front half-ring structure of the second-order BIS $\mathcal{B}_2$ holds $h_2(\tilde{\mathbf{k}})<0$. Therefore, the 3D winding number is given by $\nu_3=\mathcal{C}_1^{(3)}=1$. Similarly, the observation of third-order topological charges are simpler to determine the bulk topology.

Particularly, here we can also identify the third-order topological charges by a minimum measurement scheme (see Appendix~\ref{appendix-3}) with advantage in future experiments.
By measuring the TASP on some 2D planes of BZ, $\overline{\langle\gamma_0(\mathbf{k})\rangle}_1=0$ is for all momentum $\mathbf{k}$ on the 2D plane of $k_x=0$ (The 2D plane of $k_x=-\pi$ is failed to identify the third-order BISs $\mathcal{B}_{3}$). Thus $\overline{\langle\gamma_{0}(\mathbf{k})\rangle}_{0,2}$ on this plane reflects the locations of the second-order BISs $\mathcal{B}_2$ and third-order BISs $\mathcal{B}_3$ [see Figs.~\ref{Fig3}(d1) and \ref{Fig3}(d2)]. Therefore, the TASP $\overline{\langle\gamma_{0}(\mathbf{k})\rangle}_{3}$ and the normalized dynamic field $\Theta (\mathbf{k})$ on 2D plane of $k_x=0$ give the properties of the third-order topological charges and the topological number of the system [see Figs.~\ref{Fig3}(d3) and \ref{Fig3}(d4)].

\begin{figure*}[!htbp]
\begin{center}
\rotatebox{0}{\resizebox {14.4cm}{8.4cm} {\includegraphics{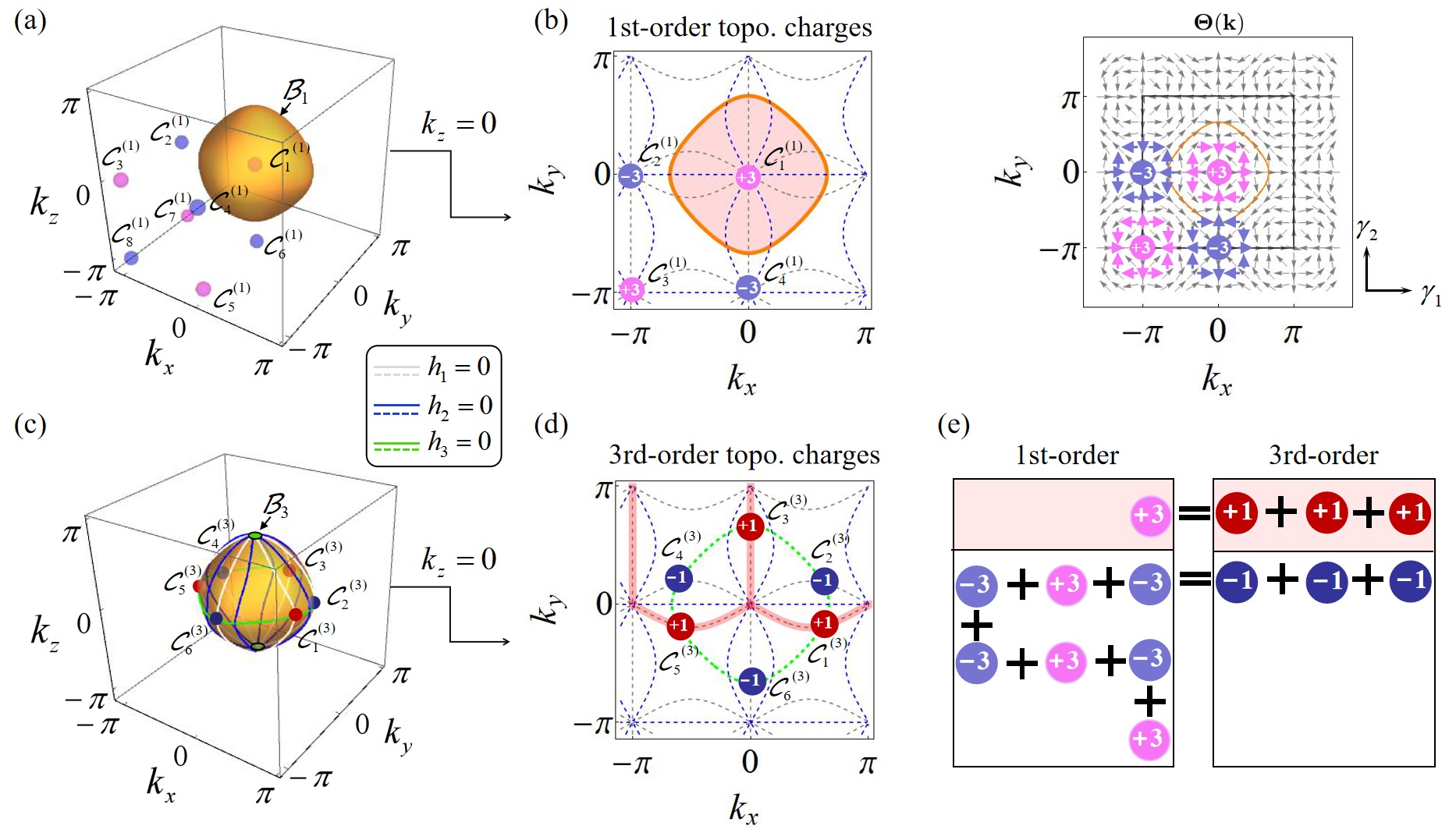}}}
\caption{Numerical results of the extended 3D chiral topological insulator model with charge value $|\mathcal{C}^{(1)}_n|=3$.
(a)~The TASP $\langle\gamma_0(\mathbf{k})\rangle_0$, where the spherical-like surface (orange surface) is identified as the first-order BIS $\mathcal{B}_1$ and eight first-order topological charges $\mathcal{C}^{(1)}_{n=1,2,\cdots,8}$ determined by $h_1(\mathbf{k})=h_2(\mathbf{k})=h_3(\mathbf{k})=0$ are marked as light-pink and light-blue points.
(b)~On 2D plane of $k_z=0$, the first-order topological charge $\mathcal{C}^{(1)}_{1}$ in the region $h_0(\mathbf{k})<0$ (light-red region) gives the 3D winding number $\nu_3=\mathcal{C}^{(1)}_{1}=3$, where the pseudospin textures in the normalized dynamic field present that the charge value of each first-order topological charge are $|\mathcal{C}^{(1)}_{n=1,2,\cdots,8}|=3$.
(c)~The second-order BISs $\mathcal{B}_2$ present ring-shape curves (gray curves) on first-order BISs $\mathcal{B}_1$ and the third-order BISs $\mathcal{B}_3$ present two points (green points) on the second-order BISs $\mathcal{B}_2$, which are
identified by the vanishing polarization $\langle\gamma_0(\mathbf{k})\rangle_{1,2}=0$. Six third-order topological charges (red and blue points) with $|\mathcal{C}^{(3)}_{n=1,2,\cdots,6}|=1$ are determined by $h_3(\tilde{\mathbf{k}})=0$ (green curve).
(d)~On 2D plane of $k_z=0$, the summation of third-order topological charges in the region $h_2(\tilde{\mathbf{k}})<0$ (light-red region) gives the 3D winding number $\nu_3=\mathcal{C}^{(3)}_{1}+\mathcal{C}^{(3)}_{3}+\mathcal{C}^{(3)}_{5}=3$.
(e)~The first-order topological charge $\mathcal{C}^{(1)}_{1}=3$ enclosed by $\mathcal{B}_1$ is equivalent to the sum of three third-order topological charges $\mathcal{C}^{(3)}_{n=1,3,5}=1$ enclosed by $\mathcal{B}_2$.
The sum of remaining first-order topological charges $\mathcal{C}^{(1)}_{n=2,\cdots,8}$ is equivalent to the sum of three third-order topological charges $\mathcal{C}^{(3)}_{n=2,4,6}=-1$. Here the other parameter is $t_{\text{so}}=t_0$.}
\label{Fig8}
\end{center}
\end{figure*}

\section{Decomposition of high integer-valued topological charges}\label{section4}

For a monopole charge without linear
dispersion, the charge value is larger than one and the system has a high-valued winding
or Chern number. If detecting the topological charges with high charge value to identify the topological phases, it is cumbersome for the measurements of the continuous charge-related (pseudo)spin texture. Nevertheless, we can avoid these redundant measurements
by reducing the high integer-valued topological charges
to multiple highest-order topological charges with unit
charge value. This essential advantage of the highest-order topological charge greatly simplifies topological characterization, especially for high-dimensional systems. Next we use two extended models to illustrate this point.

We first extend the 2D QAH model as follows:
\begin{equation}
\begin{split}
&\mathcal{H}_{\text{2D}}(\mathbf{k})=h_0\sigma_z+h_1\sigma_y+h_2\sigma_x,\\
&h_0=m_0-t_0(\cos k_x + \cos k_y),\\
&h_1=m_1+t_{\text{so}} \Im [(\sin k_x+\mathbbm{i} \sin k_y)^p],\\
&h_2=m_2+t_{\text{so}} \Re [(\sin k_x+\mathbbm{i} \sin k_y)^p],\\
\end{split}
\end{equation}
with positive integer $p$. For $m_1=m_2=0$, the topological phase corresponds to $0<|m_0|<2t_0$ with $\text{Ch}_1=-p\times\text{sgn}(m_0)$, but the trivial phase is still for $|m_0|\geqslant 2t_0$.
By quenching the system with $p=2$ in the same parameters as 2D QAH model, a ring-shape structure is identified as the first-order BIS $\mathcal{B}_1$ from the vanishing polarization $\overline{\langle\sigma_z(\mathbf{k}) \rangle}_{z}=0$ [see Fig.~\ref{Fig6}(a)]. Further, four first-order topological charges $\mathcal{C}^{(1)}_{n=1,2,3,4}$ with high charge value $|\mathcal{C}^{(1)}_{n}|=2$ are given by $\overline{\langle\sigma_z(\mathbf{k}) \rangle}_{y}=\overline{\langle\sigma_z(\mathbf{k}) \rangle}_{x}=0$ [see Fig.~\ref{Fig6}(b)].
Thus the bulk topology is determined by the summation of the first-order topological
charges enclosed by the first-order BIS $\mathcal{B}_1$, i.e. $\text{Ch}_1=\mathcal{C}^{(1)}_{2}=-2$ [see Fig.~\ref{Fig6}(d)].

After taking $\mathbf{h}^{(1)}_{\text{so}}(\tilde{\mathbf{k}})=h_2$ for the dimension reduction, we observe the second-order BISs $\mathcal{B}_2$ from the vanishing polarization $\overline{\langle\sigma_z(\mathbf{k}) \rangle}_{y}=0$, which is confined on the first-order BIS $\mathcal{B}_1$ [see Fig.~\ref{Fig6}(b)]. Correspondingly, four second-order topological charges $\mathcal{C}^{(2)}_{n=1,2,3,4}$ are obtained by the vanishing polarization $\overline{\langle\sigma_z(\mathbf{k}) \rangle}_{x}=0$ [see Fig.~\ref{Fig6}(c)]. We emphasize that now each second-order topological charge has unit charge value $|\mathcal{C}^{(2)}_n|=1$, and then the bulk topology is calculated by the summation of second-order topological charges enclosed by the second-order BISs $\mathcal{B}_2$, i.e. $\text{Ch}_1=\mathcal{C}^{(2)}_{1}+\mathcal{C}^{(2)}_{3}=-1-1=-2$ [see Fig.~\ref{Fig6}(e)]. One can find that a first-order topological charge $\mathcal{C}^{(1)}_{2}$ is separated into two second-order topological charges $\mathcal{C}^{(2)}_{n=1,3}$ with unit negative charge by dimension reduction,
and the total contribution of the remaining first-order topological charges $\mathcal{C}^{(1)}_{n=1,3,4}$ is equivalent to two second-order topological charges $\mathcal{C}^{(2)}_{n=2,4}$ with unit positive charge [see Fig.~\ref{Fig6}(f)]. The 2D topology with high integer-valued $\text{Ch}_1=2$ is transformed to 0D topology given by the summation of two 0th Chern numbers with $\text{Ch}_0=1$, i.e. $\text{Ch}_1\rightarrowtail 2\text{Ch}_0$.

Similarly, we extend the 3D chiral topological insulator model as the following case,
\begin{align}
\mathcal{H}_{\text{3D}}(\mathbf{k})&=h_0 \sigma_z\otimes \tau_x+h_1\sigma_x\otimes \mathbbm{1}+h_2\sigma_y\otimes \mathbbm{1}+h_3\sigma_z\otimes \tau_z,\notag\\
&h_0=m_0-t_0(\cos k_x + \cos k_y + \cos k_z),\notag\\
&h_1=m_1+t_{\text{so}}( \sin ^3 k_x-3\sin k_x \sin ^2 k_y),\notag\\
&h_2=m_2+t_{\text{so}}( 3\sin k_y \sin ^2 k_x -\sin ^3 k_y),\notag\\
&h_3=m_3+t_{\text{so}} \sin ^3 k_z \label{3ed}.
\end{align}
For $m_1=m_2=m_3=0$, the topological phases are classified by: (i) $t_0<m_0<3t_0$ with $\nu_3=3$; (ii) $-t_0<m_0<t_0$ with $\nu_3=-6$; and (iii) $-3t_0<m_0<-t_0$ with $\nu_3=3$. By taking the quenched parameters as the same as 3D chiral topological insulator model, eight first-order topological charges with high charge value $|\mathcal{C}^{(1)}_{n=1,2,\cdots,8}|=3$ and a spherical-like first-order BIS $\mathcal{B}_1$ [see Figs.~\ref{Fig8}(a) and \ref{Fig8}(b)] can be observed by vanishing polarization of $\langle\gamma_0(\mathbf{k})\rangle_{0,1,2,3}$ when taking $\mathbf{h}^{(0)}_{\text{so}}(\mathbf{k})=(h_1,h_2,h_3)$. Thus the bulk topology is determined by the summation of first-order topological charges enclosed by the first-order BISs $\mathcal{B}_1$, i.e. $\nu_3=\mathcal{C}^{(1)}_{1}=3$.

After taking $\mathbf{h}^{(2)}_{\text{so}}(\tilde{\mathbf{k}})=h_3$ for the dimension reduction, six third-order topological charges $\mathcal{C}^{(3)}_{n=1,2,\cdots,6}$ sit on the second-order BISs $\mathcal{B}_2$ [see Fig.~\ref{Fig8}(c)]. Each third-order topological charge has unit charge value $|\mathcal{C}^{(3)}_{n=1,2,\cdots,6}|=1$, and then the bulk topology is given by the summation of third-order topological charges enclosed by the third-order BISs $\mathcal{B}_3$, i.e. $\nu_3=\mathcal{C}^{(3)}_{1}+\mathcal{C}^{(3)}_{3}+\mathcal{C}^{(3)}_{5}=3$ [see Fig.~\ref{Fig8}(d)]. Similarly, a first-order topological charge $\mathcal{C}^{(1)}_{1}$ is separated into three third-order topological charges $\mathcal{C}^{(3)}_{n=1,3,5}$ with unit positive charge,
and the total contribution of the remaining first-order topological charges $\mathcal{C}^{(1)}_{n=2,\cdots ,8}$ is equivalent to the summation of three third-order topological charges $\mathcal{C}^{(3)}_{n=2,4,6}$ with unit negative charge [see Fig.~\ref{Fig8}(e)].
Thus a high integer-valued $3$D winding number with $\nu_3=3$ is transformed to the sum of three $0$th Chern numbers, i.e. $\nu_3\rightarrowtail 3\text{Ch}_0$.

The above results strongly demonstrate the advantages of the highest-order topological charge in characterization of topological phases. Although the definition of topological charge depends on the selection of the $h$-components, choosing a different  $h$-component to define the highest-order topological charge will not change the essence of its unit charge value, which is different from the first-order topological charge.
Therefore, for a more general system, we only need to measure the properties of the highest-order topological charge to determine the bulk topology.

\section{Conclusion and discussion}

In conclusion, we have proposed a new dynamical scheme to characterize the equilibrium topological phases based on the high-order topological charges, which correspond to monopoles confined in low dimensional subspaces. Through a dimensional reduction approach for a $d$D bulk Hamiltonian, the topology of the $d$D system can be determined by the arbitrary $s$th order topological charges enclosed by the $s$th-order BISs. In quenching the system from a trivial phase to a topologically nontrivial regime, both the high-order BISs and the high-order topological charges are directly observed by the quench induced (pseudo)spin dynamics, for which the topological phases of post-quench Hamiltonian can be detected dynamically.

The high-order topological charges have essential advantages in characterizing topological phases due to their intrinsic features. We compare the first-order and highest-order topological charges. For the first-order topological charge with unit or high charge value, as defined in $d$D momentum space, its characterization generically necessitates to measure the continuous charge-related (pseudo)spin texture in $d$D space. In comparison, the highest-order topological charges are defined in the zero dimension, and are characterized by the discrete signs of spin-polarization in zero dimension. This intrinsic feature determines that the charge value of a highest-order topological charge only takes $\mathcal{C}^{(d)}=\pm1$. Accordingly, a high integer-valued lower-order topological charge can always reduce to multiple highest-order topological charges with unit charge value, which can be easily measured in experiment, hence simplifying the characterization and detection of topological phases.

\section*{ACKNOWLEDGEMENT}
\par This work was supported by National Natural Science Foundation of China (Grants No. 11761161003, No. 11825401, and No. 11921005), the National Key R\& D Program of China (Project No. 2016YFA0301604), Strategic Priority Research Program of the Chinese Academy of Science (Grant No. XDB28000000), and by the Open Project of Shenzhen Institute of Quantum Science and Engineering (Grant No.SIQSE202003).

\appendix

\section{Deep quench process}\label{appendix-2}

In quenching the axis $\gamma_i$, we initialize a fully polarized state $\rho_i(0)$ along the opposite $\gamma_i$ axis by introducing a very large constant magnetization $m_i$ such that $h_i(\mathbf{k})\approx m_i\gg 0$ for $t<0$. After $t=0$, the magnetization $m_i$ is suddenly tuned to the topological regime, and the momentum-linked (pseudo)spin expectation $\langle \boldsymbol{\gamma}(\mathbf{k},t)\rangle$ will process around $\mathbf{h}(\mathbf{k})$. The quantum dynamics is governed by the unitary evolution operator $\mathcal{U}(t)=\text{exp}(-\mathbbm{i}\mathcal{H}t)$ with the post-quenched Hamiltonian $\mathcal{H}(\mathbf{k})$. We can measure the time-averaged (pseudo)spin polarization (TASP) of the component $\gamma_0$,
\begin{align}
    \overline{\langle\gamma_{0}(\mathbf{k})\rangle}_{i} & \equiv\lim_{T\to\infty}\frac{1}{T}\int_{0}^{T}\mathrm{d}t\,\mathrm{Tr}[\rho_{i}(0)e^{\mathrm{i}\mathcal{H}(\mathbf{k})t}\gamma_{0}e^{-\mathrm{i}\mathcal{H}(\mathbf{k})t}]\nonumber \\
     & =-h_{0}(\mathbf{k})h_{i}(\mathbf{k})/E^{2}(\mathbf{k}),\label{eb}
\end{align}
where $E(\mathbf{k})=\sqrt{\sum^d_{i=0} h^2_i}$ is the energy of the post-quenched Hamiltonian.

\section{Minimal measurement scheme}\label{appendix-3}

We provide a minimal dynamical scheme for the topological systems, in which the bulk topology is determined by the $d$th-order topological charges $\mathcal{C}_n^{(d)}$ enclosed by the $0$D $d$th-order BISs $\mathcal{B}_d$. This scheme greatly simplifies the characterization of bulk topological phases, especially for $d\geqslant 3$. We consider the topological systems with at least one plane-type component, say $h_{d-2}$, which means that the momenta satisfying $h_{d-2}(\mathbf{k})=0$ form planes. Note that both the $d$th-order topological charges and the $d$th-order BISs sit on the $1$D $(d-1)$th-order BISs $\mathcal{B}_{d-1}$ consisting of momenta with $h_{0}=h_{1}=\cdots=h_{d-2}=0$. Since $h_{d-2}$ is plane-type, the $(d-1)$th-order BISs $\mathcal{B}_{d-1}$ also belong to the plane determined by $h_{d-2}=0$. With these observations, to identify the $d$th-order topological charges and the corresponding BISs, we can first extract the planes specified by $h_{d-2}$ from the TASP $\overline{\langle \gamma_0(\mathbf{k}) \rangle}_{d-2}$ with vanishing values. On these planes, $\overline{\langle \gamma_0(\mathbf{k})\rangle}_{0}=\overline{\langle \gamma_0(\mathbf{k})\rangle}_{1}=\cdots=\overline{\langle \gamma_0(\mathbf{k})\rangle}_{d-3}=0$ further determine the $(d-1)$th order BISs $\mathcal{B}_{d-1}$. Finally, the $d$th-order BISs $\mathcal{B}_{d}$ and the $d$th-order topological charges shall be found by observing $\overline{\langle \gamma_0(\mathbf{k})\rangle}_{d-1}=0$ and $\overline{\langle \gamma_0(\mathbf{k})\rangle}_{d}=0$ in the $(d-1)$th BISs.

\begin{figure}[!htbp]
\begin{center}
\rotatebox{0}{\resizebox {8.0cm}{7.2cm} {\includegraphics{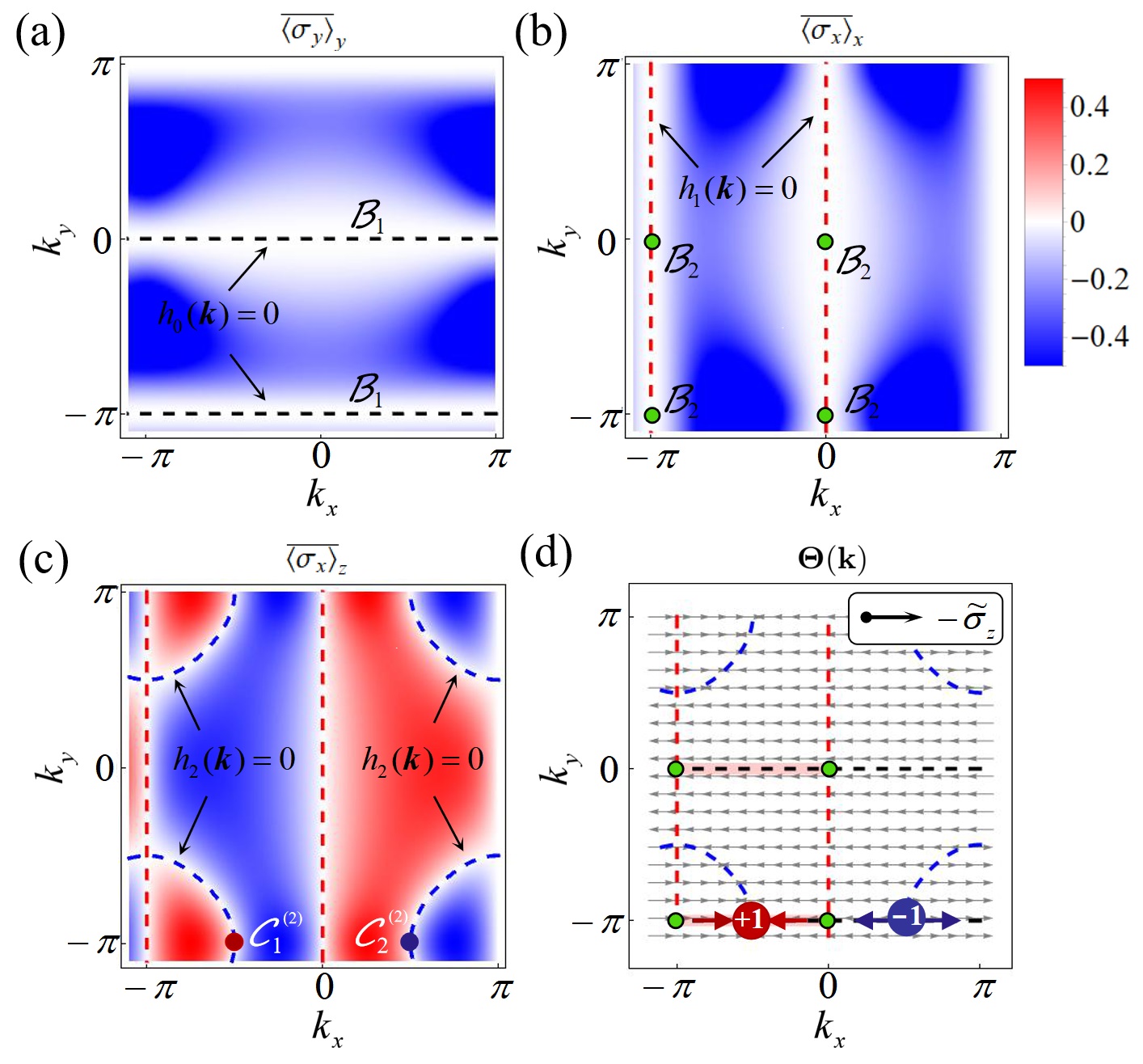}}}
\caption{Dynamical characterization of 2D QAH model. (a)-(c)~The TASP $\overline{\langle\sigma_y(\mathbf{k}) \rangle}_{y}$
and $\overline{\langle\sigma_x(\mathbf{k}) \rangle}_{x,z}$, where the vanishing polarization marked as the black, red, and blue dashed line presents the interface with $h_{0,1,2}(\mathbf{k})=0$, respectively. The first-order BISs $\mathcal{B}_1$ (two black dashed line) are identified by $\overline{\langle\sigma_y(\mathbf{k}) \rangle}_{y}=0$ in (a). The second-order BISs $\mathcal{B}_2$ are four points (green) at $(-\pi,-\pi)$, $(-\pi,0)$, $(0,-\pi)$, and $(0,0)$, which are given by $\overline{\langle\sigma_x(\mathbf{k}) \rangle}_{x}=0$ on the first-order BISs $\mathcal{B}_1$ in (b).
The second-order topological charges $\mathcal{C}_1^{(2)}$ and $\mathcal{C}_2^{(2)}$ at $(-\pi/2,-\pi)$ and $(\pi/2,-\pi)$ are determined by $h_2(\tilde{\mathbf{k}})=0$ of $\overline{\langle\sigma_x(\mathbf{k}) \rangle}_{z}$ on the first-order BISs $\mathcal{B}_1$ in (c).
(d)~The normalized dynamic field in $\tilde{\sigma}_z$ spin subspace characterize the properties of the topological charges, where $\mathcal{C}_1^{(2)}=1$ in the region $h_1(\tilde{\mathbf{k}})<0$ (light-red thick-solid curves) is enclosed by the second-order BISs $\mathcal{B}_2$ and gives the Chern number $\text{Ch}_1=\mathcal{C}_1^{(2)}=1$. Here the other parameter is $t_{\text{so}}=t_0$.}
\label{Figs1}
\end{center}
\end{figure}

\begin{figure*}[htbp]
\centering
\rotatebox{0}{\resizebox {14.4cm}{7.4cm} {\includegraphics{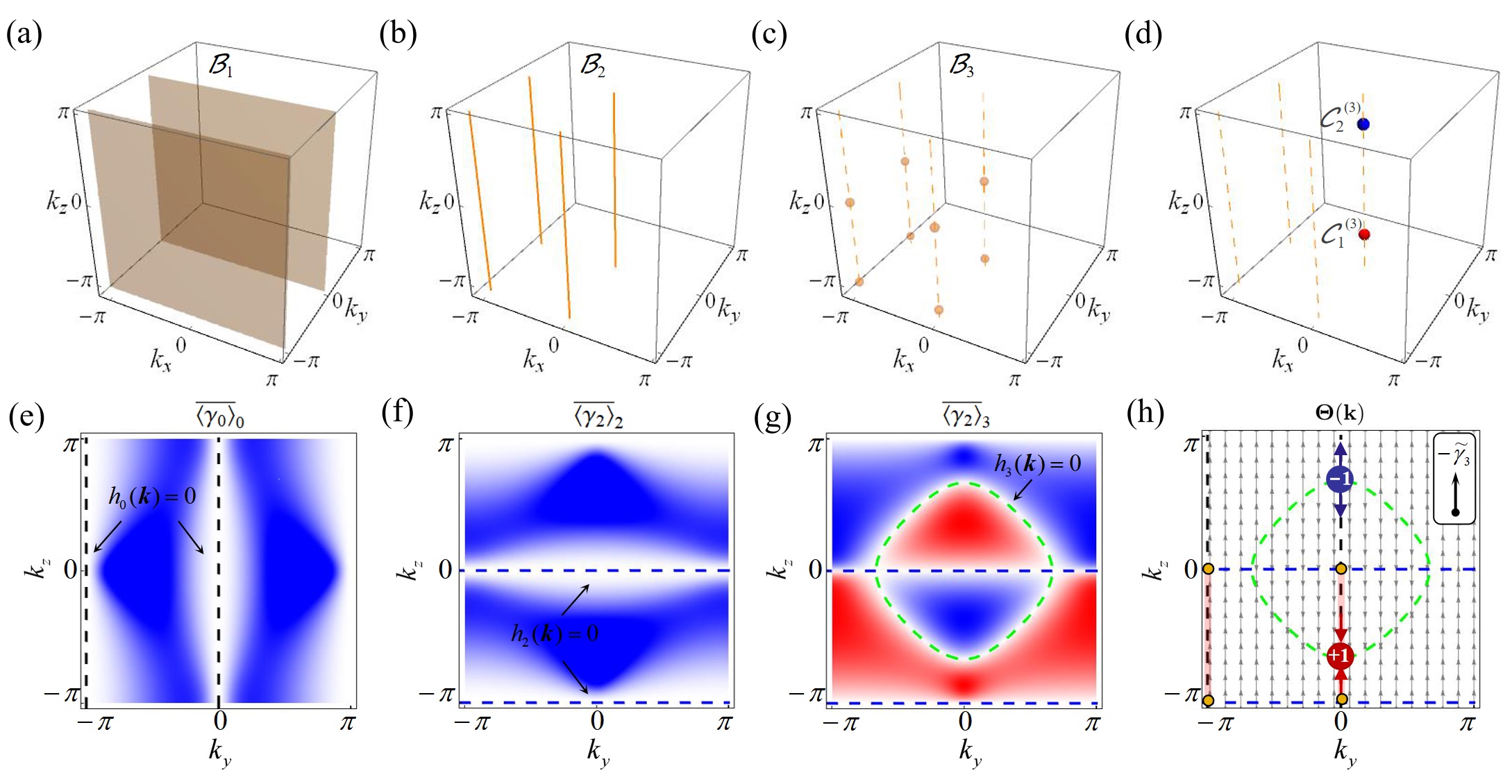}}}
\caption{Dynamical characterization of 3D chiral topological insulator. (a-d)~TASP via quenching $(m_0,m_1,m_2,m_3)$, where two planes (orange surface) of $k_y=0$ and $k_y=-\pi$ present $\overline{\langle\gamma_0(\mathbf{k}) \rangle}_{0}=0$ which are identified the first-order BISs $\mathcal{B}_1$ with $h_{0}(\mathbf{k})=0$ in (a). Four lines of $\overline{\langle\gamma_1(\mathbf{k}) \rangle}_{1}=0$ on the first-order BISs $\mathcal{B}_1$ at $k_x=0$ and $k_x=-\pi$ are the second-order BISs $\mathcal{B}_2$ (orange lines) in (b).
The third-order BISs $\mathcal{B}_3$ present eight points given by $\overline{\langle\gamma_2(\mathbf{k}) \rangle}_{2}=0$ on the second-order BISs $\mathcal{B}_2$ (orange points) in (c).
Two third-order topological charges $\mathcal{C}_1^{(3)}$ (red point) and $\mathcal{C}_2^{(3)}$ (blue point) at $(k_x,k_y,k_z)=(0,0,-2\pi/3)$ and $(0,0,2\pi/3)$ are determined by $h_3(\tilde{\mathbf{k}})=0$ of $\overline{\langle\gamma_2(\mathbf{k}) \rangle}_{3}$ on the second-order BISs $\mathcal{B}_2$ in(d), which gives $\nu_3=\mathcal{C}_1^{(3)}=1$.
(e-h) Minimal measurement for the TASP in $k_x=0$, where $\overline{\langle\gamma_0(\mathbf{k}) \rangle}_{0}$, $\overline{\langle\gamma_2(\mathbf{k}) \rangle}_{2}$, and $\overline{\langle\gamma_2(\mathbf{k}) \rangle}_{3}$ in (e), (f), and (g). The vanishing polarization marked as the black, blue and green dashed line presents the interface with $h_{0(1),2,3}(\mathbf{k})=0$, respectively. The normalized dynamic field characterizes the properties of the third-order topological charges in (h), where $\mathcal{C}_1^{(3)}=1$ in the region $h_2(\tilde{\mathbf{k}})<0$ (light-red thick-solid curves) is enclosed by $\mathcal{B}_3$ and gives the winding number $\nu_3=1$. Here the other parameter is $t_{\text{so}}=t_0$.}
\label{Fig-s2}
\end{figure*}

\section{Another dynamical characterization scheme}\label{appendix-4}

We provide another dynamical characterization scheme by quenching all (pseudo)spin axes and measuring multiple (pseudo)spin axis. On the other hand, we notice that the configurations of the high-order BISs and high-order topological charges are sharply different if choosing different components ($h_i$) of the Hamiltonian for definition. Here we take the different $h$-components to define the high-order topological charges compared with the previous results.

We quench $s$ axes and measure the same axes for determination of $s$th-order high-order BISs through TASP,
\begin{equation}
\mathcal{B}_s={\{\mathbf{k}\in\mathrm{BZ}\vert\overline{\langle\gamma_0\rangle}_0=\cdots =\overline{\langle\gamma_{s-1}\rangle}_{s-1}=0\}},
\end{equation}
and then the $s$th-order topological charges are identified by quenching the remaining
axes and only measuring the $\gamma_{s-1}$ component, i.e. $\overline{\langle\gamma_{s-1}(\mathbf{k})\rangle}_{j}$ with $j=s,s+1,\cdots,d$. We further define
\begin{equation}\label{eg2}
\Theta_j(\tilde{\mathbf{k}})\equiv -\lim_{\mathbf{k}\to\tilde{{\mathbf{k}}}}\frac{\text{sgn}[h_{s-1}({\mathbf{k}})]}{\mathcal{N}_{{\mathbf{k}}}}\overline{\langle\gamma_{s-1}(\mathbf{k})\rangle}_{j}
\end{equation}
in (pseudo)spin subspace with the coordinate system $\tilde{\gamma}_{s}\text{-}\tilde{\gamma}_{s+1}\text{-}\cdots\text{-}\tilde{\gamma}_{d}$, where $\mathcal{N}_{\tilde{\mathbf{k}}}$ is a normalization factor. Near the monopole charge, the dynamic field satisfies $\Theta_j(\tilde{\mathbf{k}})|_{\tilde{\mathbf{k}}\rightarrow \mathfrak{g}_n}=h^{(s-1)}_{\text{so},j}(\tilde{\mathbf{k}})$, thus the high-order topological charge is determined directly by
$ \mathcal{C}^{(s)}_n=\text{sgn}[J_{\boldsymbol{\Theta}}(\mathfrak{g}_n)]$
in the linear case. Note that the above dynamical characterization scheme is same with that in previous results for the determination of first-order topological charges.
We next numerically examine the 2D QAH model and 3D chiral topological insulator model, and only consider the highest-order cases.

For 2D QAH model $\mathcal{H}_{\text{2D}}(\mathbf{k})=h_x(\mathbf{k})\sigma_x+h_y(\mathbf{k})\sigma_y+h_z(\mathbf{k})\sigma_z$, we reselect
\begin{equation}
\begin{split}
&h_0=h_y=m_y+t_\text{so}\sin k_y,\\
&h_1=h_x=m_x+t_\text{so}\sin k_x,\\
h_2=&h_z=m_z-t_0\cos k_x-t_0\cos k_y.\\
\end{split}
\end{equation}
By quenching $(m_y,m_x,m_z)$ from $(30t_0,0,0)$ to $(0,0,-t_0)$ for $h_{y}$ and measuring the spin polarization of $\gamma_y$-component, the TASP $\overline{\langle\sigma_y(\mathbf{k}) \rangle}_{y}$ is obtained. We observe that the first-order BISs $\mathcal{B}_1$ are identified by $\overline{\langle\sigma_y(\mathbf{k}) \rangle}_{y}=0$, which are two lines in Fig.~\ref{Figs1}(a). Further, we quench the $(m_y,m_x,m_z)$ of system from $(0,30t_0,0)$ to $(0,0,-t_0)$ for $h_{x}$ and from $(0,0,30t_0)$ to $(0,0,-t_0)$ for $h_{z}$. After only measuring the spin polarization of $\gamma_x$-component,
the TASP $\overline{\langle\sigma_x(\mathbf{k}) \rangle}_{x,z}$ are obtained. We observe that the second-order BISs $\mathcal{B}_2$ present four points at $(-\pi,-\pi)$, $(-\pi,0)$, $(0,-\pi)$, and $(0,0)$, which are given by $\overline{\langle\sigma_x(\mathbf{k}) \rangle}_{x}=0$ on the first-order BISs $\mathcal{B}_1$ in Fig.~\ref{Figs1}(b). Finally, two second-order topological charges $\mathcal{C}_1^{(2)}$ and $\mathcal{C}_2^{(2)}$ at $(-\pi/2,-\pi)$ and $(\pi/2,-\pi)$ are determined by $h_2(\tilde{\mathbf{k}})=0$ of $\overline{\langle\sigma_x(\mathbf{k}) \rangle}_{z}$ on the first-order BISs $\mathcal{B}_1$, as shown in Fig.~\ref{Figs1}(c). The bulk topology is determined by $\mathcal{C}_1^{(2)}=1$ enclosed by the second-order BISs $\mathcal{B}_2$, i.e. $\text{Ch}_1=\mathcal{C}_1^{(2)}=1$.

We further consider the 3D chiral topological insulator model $\mathcal{H}_\text{3D}(\mathbf{k})=\sum^3_{i=0} h_i(\mathbf{k})\gamma_i$ and
reselect the component $h_i$ as follows:
\begin{equation}
\begin{split}
&h_0=m_2+t_{\text{so}}\sin k_y,\\
&h_1=m_1+t_{\text{so}}\sin k_x,\\
&h_2=m_3+t_{\text{so}}\sin k_z,\\
h_3=&m_0-t_0(\cos k_x +\cos k_y +\cos k_z),\\
\end{split}
\end{equation}
where the $\boldsymbol{\gamma}$ matrices are taken as $\gamma_0=\sigma_y\otimes \mathbbm{1}$, $\gamma_1=\sigma_x\otimes \mathbbm{1}$, $\gamma_2=\sigma_z\otimes \tau_z$, and $\gamma_3=\sigma_z\otimes \tau_x$.
When the quench is firstly performed by suddenly varying $(m_0,m_1,m_2,m_3)$ from $(0,0,30t_0,0)$ to $(1.5t_0,0,0,0)$ for $h_0$ and then the pseudospin polarization of $\gamma_0$-component is measured,
we observe that the first-order BISs $\mathcal{B}_1$ are identified by $\overline{\langle\gamma_0(\mathbf{k}) \rangle}_{0}=0$, which are two planes of $k_y=0$ and $k_y=-\pi$, as shown in Fig.~\ref{Fig-s2}(a).
Secondly, we quench $(m_0,m_1,m_2,m_3)$ from $(0,30t_0,0,0)$ to $(1.5t_0,0,0,0)$ for $h_1$ and measure the pseudospin polarization of $\gamma_1$-component, the second-order BISs $\mathcal{B}_2$ are identified by $\overline{\langle\gamma_1(\mathbf{k}) \rangle}_{1}=0$, which are four lines on the first-order BISs $\mathcal{B}_1$ at $k_x=0$ and $k_x=-\pi$, as shown in Fig.~\ref{Fig-s2}(b).
Thirdly, we quench $(m_0,m_1,m_2,m_3)$ from $(0,0,0,30t_0)$ to $(1.5t_0,0,0,0)$ for  $h_2$ and from $(0,0,0,0)$ to $(1.5t_0,0,0,0)$ for $h_3$.
By only measuring the pseudospin polarization of $\gamma_2$-component, the TASP $\overline{\langle\gamma_{2}(\mathbf{k}) \rangle}_{2,3}$ are obtained. We observe that the third-order BISs $\mathcal{B}_3$ present eight points given by $\overline{\langle\gamma_2(\mathbf{k}) \rangle}_{2}=0$ on the second-order BISs $\mathcal{B}_2$, as shown in Fig.~\ref{Fig-s2}(c). Finally, two third-order topological charges $\mathcal{C}_1^{(3)}$ and $\mathcal{C}_2^{(3)}$ at $(k_x,k_y,k_z)=(0,0,-2\pi/3)$ and $(0,0,2\pi/3)$ are determined by $h_3(\tilde{\mathbf{k}})=0$ of $\overline{\langle\gamma_2(\mathbf{k}) \rangle}_{3}$ on the second-order BISs $\mathcal{B}_2$, as shown in Fig.~\ref{Fig-s2}(d). Thus the bulk topology is determined by $\mathcal{C}_1^{(3)}=1$ enclosed by the third-order BISs $\mathcal{B}_3$, i.e. $\nu_3=\mathcal{C}_1^{(3)}=1$.

Besides, we also give the 2D measurement to determine the bulk topology based on the minimal scheme of Appendix \ref{appendix-3}. For this 3D model, the second-order BISs $\mathcal{B}_2$ must be on 2D planes. One can measure the TASP $\overline{\langle\gamma_1(\mathbf{k}) \rangle}_{1}$ on 2D planes after quench, then $\overline{\langle\gamma_1(\mathbf{k}) \rangle}_{1}=0$ is for all $\mathbf{k}$ on 2D planes of $k_x=0$ and $k_x=-\pi$. Therefore, the third-order BISs $\mathcal{B}_3$ and the corresponding third-order topological charges are obtained by the TASP $\overline{\langle\gamma_0(\mathbf{k}) \rangle}_{0}$ and $\overline{\langle\gamma_2(\mathbf{k}) \rangle}_{2,3}$, as shown in Figs.~\ref{Fig-s2}(e-g). The bulk topology is characterized by $\mathcal{C}_1^{(3)}=1$, as shown in Fig.~\ref{Fig-s2}(h).

\bibliography{ref}

\end{document}